\documentclass[useAMS,usenatbib]{mn2e}
\usepackage{graphicx}
\usepackage{epstopdf}
\usepackage{ulem}

\title[Identification of Globular Cluster Stars in RAVE data II: Tidal debris around NGC 3201]{Identification of Globular Cluster Stars in RAVE data II: Extended tidal debris around NGC 3201}
\author[B. Anguiano et al.]{B.~Anguiano$^{1}$\thanks{E-mail: borja.anguiano@mq.edu.au}, G.~M.~De Silva$^{2,3}$, K.~Freeman$^{4}$, G.~S.~Da Costa$^{4}$, T.~Zwitter$^{5}$, \newauthor A.~C.~Quillen$^{6}$, D. B.~Zucker$^{1,2}$, J. F.~Navarro$^{7}$, A.~Kunder$^{8}$, A.~Siebert$^{9}$, R.~F.~G.~Wyse$^{10}$, \newauthor E.~K.~Grebel$^{11}$, G.~Kordopatis$^{8}$, B. K~Gibson$^{12}$, G.~Seabroke$^{13}$, S.~Sharma$^{3}$, J.~Wojno$^{8}$\newauthor J.~Bland-Hawthorn$^{3}$, Q.~A.~Parker$^{2}$, M.~Steinmetz$^{8}$, C.~Boeche$^{11}$, G.~Gilmore$^{14}$, \newauthor  O.~Bienaym\'e$^{9}$, W.~Reid$^{1}$, F.~Watson$^{2}$ \\
\\
$^{1}$ Department of Physics and Astronomy, Macquarie University, North Ryde, NSW 2109, Australia\\
$^{2}$ Australian Astronomical Observatory, North Ryde, NSW 2113, Australia\\
$^{3}$ Sydney Institute for Astronomy, School of Physics, The University of Sydney, NSW 2006, Australia\\
$^{4}$ Research School of Astronomy and Astrophysics, Australian National University, Cotter Rd., Weston, ACT 2611, Australia\\
$^{5}$ University of Ljubljana, Faculty of Mathematics and Physics, Ljubljana, Slovenia\\
$^{6}$ Department of Physics and Astronomy, University of Rochester, Rochester, NY 14627, USA\\
$^{7}$ Department of Physics and Astronomy, University of Victoria, Victoria, BC V8P 5C2, Canada\\
$^{8}$ Leibniz-Institut f\"ur Astrophysik Potsdam (AIP), An der Sternwarte 16, D-14482 Potsdam, Germany\\
$^{9}$ Observatoire astronomique de Strasbourg, Universit\'e de Strasbourg, CNRS, UMR 7550, 11 rue de l'Universit\'e, F-67000 Strasbourg, France \\
$^{10}$ Department of Physics and Astronomy, Johns Hopkins University, 3400 N. Charles St, Baltimore, MD 21218, USA\\
$^{11}$ Astronomisches Rechen-Institut, Zentrum f\"ur Astronomie der Universit\"at Heidelberg, M\"onchhofstr.\ 12--14, 69120 Heidelberg, Germany\\
$^{12}$ E. A. Milne Centre for Astrophysics, University of Hull, Cottingham Road, Hull, HU6 7RX, UK\\
$^{13}$ Mullard Space Science Laboratory, University College London, Holmbury St Mary, Dorking, RH5 6NT, UK\\
$^{14}$ Institute of Astronomy, University of Cambridge, Madingley Road, Cambridge CB3 0HA, UK}
\begin{document}

\date{}

\pagerange{\pageref{firstpage}--\pageref{lastpage}} \pubyear{2014}

\maketitle

\label{firstpage}

\begin{abstract}
We report the identification of extended tidal debris potentially associated with the globular cluster NGC 3201, using the RAVE catalogue. We find the debris stars are located at a distance range of 1-7 kpc based on the forthcoming RAVE distance estimates. The derived space velocities and integrals of motion show interesting connections to NGC 3201, modulo uncertainties in the proper motions. Three stars, which are among the 4 most likely candidates for NGC 3201 tidal debris, are separated by 80$^{\circ}$ on the sky yet are well matched by the 12 Gyr, [Fe/H] = -1.5 isochrone appropriate for the cluster.  This is the first time tidal debris around this cluster has been reported over such a large spatial extent, with implications for the clusterÕs origin and dynamical evolution.
\end{abstract}

\begin{keywords}
Galaxy: kinematics and dynamics -- (Galaxy): globular clusters: individual: NGC 3201
\end{keywords}

\section{Introduction}

Tidal tails or debris associated with globular clusters (GCs) are important substructures that carry significant information on the dynamical history of the clusters and their host galaxies. All Galactic GCs are expected to have lost mass through the escape of individual stars \citep[e.g.][]{1995AJ....109.2553G,1999A&A...352..149C, 2000A&A...359..907L, 2009ApJ...693.1118G,2010A&A...522A..71J,2014A&A...572A..30K,2015A&A...574A..15F}. See also \citet{2003MNRAS.340..227B} for a broad theoretical framework. The most prominent example of mass loss through individual stars is the GC Pal 5 with its associated tidal tails \citep[e.g.][]{2001ApJ...548L.165O,2002AJ....124..349R}, a cluster that is currently undergoing disruption \citep{2004AJ....127.2753D}. While tidal tails are a direct sign of mass loss, further evidence can be found in the cluster present day mass function (PDMF). Clusters showing a PDMF that is flat or inverted suggests that those objects have lost up to 90$\%$ of their initial mass \citep{2015MNRAS.453.3278W}. Note however that tails are not observed in all of these inverted PDMF clusters. Moreover, \citet{2010MNRAS.401..105K} found that the time variation of the tidal field need not lead to detectable overdensities within the tidal tails as suggested by \citet{2004AJ....127.2753D}.

Using CN and CH as tracers of stars born in GCs, \citet{2010A&A...519A..14M} concluded that 17$\%$ of the present-day stellar mass of the Galactic halo originated from dissolved GCs. \citet{2010ApJ...718L.112V} used hydrodynamical simulations to estimate the fraction of the mass of the Galactic stellar halo composed of second-generation stars that originated in GCs. They found that the fraction is always $<$ 8$\%$. There is evidence that the most luminous GCs in the Galaxy are the remnant nuclei of accreted dwarf galaxies \citep[e.g.][]{1995AJ....109.1086S, 2003MNRAS.346L..11B}. Such cases could be identified by the presence of multiple populations and/or a spread in metallicity within the cluster and debris stars. Therefore the metallicity spread could help to distinguish between a GC and a dwarf galaxy. Note that the origins of multiple populations in GCs are unclear to date, but they are not thought to be due to real age spreads \citep{2004ARA&A..42..385G}. In dwarf galaxies however the multiple populations reflect a true difference in stellar ages \citep{2012AJ....144...76W}. During the process of accreting dwarf galaxies, GCs that formed within dwarf galaxies also are accreted into the Galactic halo, e.g. the accretion of the Sagittarius dwarf galaxy, where several GCs within the dwarf galaxy are in the process of joining the Galactic GC population \citep{1995AJ....109.2533D,2010ApJ...718.1128L}. Further, tidal tails are useful for constraining the global mass of the Galaxy and its lumpiness. GC tidal tails can be some of the coldest and longest-lived substructures in the Milky Way halo \citep{1999AJ....118..911M,2009ApJ...705L.223C}. As stars in tidal tails are usually aligned with the orbit of the GC, their identification allows a better handle on modeling the cluster orbit, which in turn places constraints on the Galactic potential, barring any major disruption events such as interactions with molecular clouds. 

In the present work we report the detection of potential extended tidal debris associated with NGC 3201, a low mass halo GC. NGC 3201 has an extreme heliocentric radial velocity (RV) of 494 km s$^{-1}$ and a highly retrograde orbit \citep{2007AJ....134..195C}, suggesting an extra-Galactic origin; an additional argument is its location on the "dwarf accretion" arm in the age-metallicity diagram of \citet{2013MNRAS.436..122L}. Despite the large metallicity spread of at least 0.4 dex reported earlier by \citet{1998AJ....116..765G} and \citet{2013ApJ...764L...7S}, the latest studies suggest that a metallicity spread within the cluster members is unlikely \citep{2015ApJ...801...69M}.

This paper is organised as follows. In Section 2 we describe the sample selection. The proper motion diagram and color-magnitude analysis are described in Section 3. In Section 4 and 5 we explore the space velocities, the integrals of motion and the metallicity distribution function of the tidal debris candidates. Finally, we present our conclusions in Section 6.


\section{Sample selection}

\begin{figure*}
  \centering  
 \includegraphics[width=2.0\columnwidth]{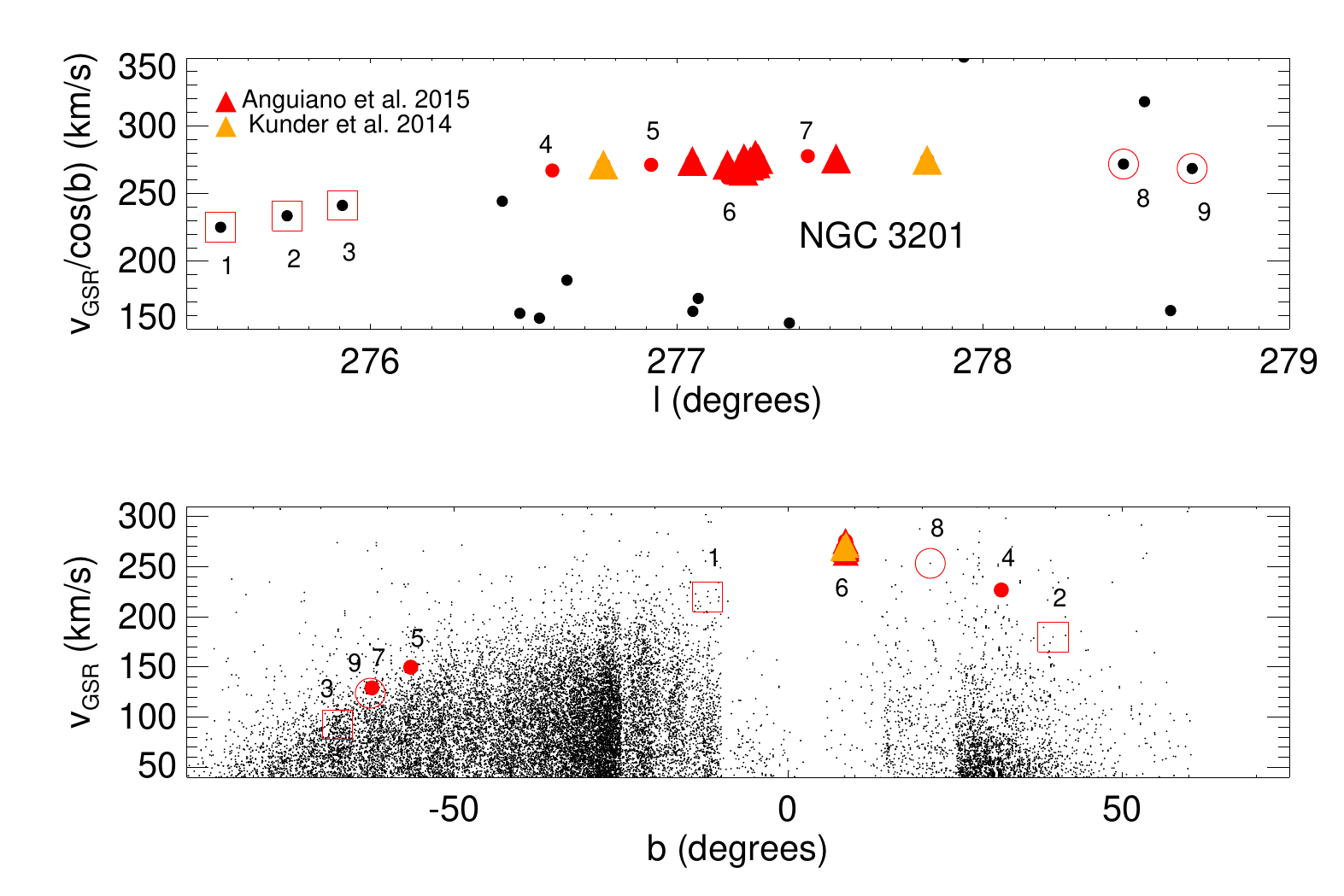}
   \caption{Radial velocity distribution in the GSR centroid at the position of NGC 3201, in the Galactic longitude range 275$^{\circ}$ $<$ l $<$ 279$^{\circ}$ (top panel) and in the latitude range -90$^{\circ}$ $<$ b $<$ 75$^{\circ}$ (bottom panel). The top panel shows the line-of-sight velocity projected to the disk plane by a $\cos$ b normalization while the bottom shows v$_{GSR}$. The numbered red open squares, open circles and red dots are the candidates of the extended tidal stream potentially associated with NGC 3201. Red and orange triangles are cluster members previously identified in \citet{2014A&A...572A..30K} and \citet{2015MNRAS.451.1229A}.}
  \label{fig:RVDF_stream}
\end{figure*}

The RAdial Velocity Experiment (RAVE) provides precise radial velocities and stellar atmosphere parameters for 483,849 objects using medium-resolution spectra (R = 7,500) in the Ca II-triplet region (8410 - 8795 \AA) \citep{2006AJ....132.1645S}. Using the fourth data release of the RAVE catalogue, RAVE--DR4 \citep{2013AJ....146..134K}, we apply the selection criteria developed in \citet{2015MNRAS.451.1229A} for data quality, where stars with a S/N $>$ 10, cross-correlation coefficient R larger than 5 and classification flag set to "n" were selected. From this sample, we further selected giant stars within the range 3600 K $<$ T$_{\rm eff}$ $<$ 6000 K, log $g$ $<$ 3.5 and [M/H] $<$ +0.0 dex. Most of the selected spectra have a S/N $>$ 20. These stars have a typical error in radial velocity $\leq$ 2 km s$^{-1}$. For this selection, we explore the radial velocity in the Galactic Standard of Rest frame (v$_{GSR}$) using the following formulae:

\begin{equation}
v_{LSR} = v_{BSR} + 9 \cos(l) \cos(b) + 12 \sin(l) \cos(b) + 7 \sin(b)
\end{equation}

\begin{equation}
v_{GSR} = v_{LSR} + 220 \sin(l) \cos(b)
\end{equation}

where LSR is the Local Standard of Rest, BSR is the Barycentric Standard of Rest\footnote{Here we make use of the heliocentric radial velocities provided by RAVE. The difference between barycentric and heliocentric velocities is negligible for our purposes.} and $l$ and $b$ are Galactic longitude and latitude respectively. Numerical coefficients are in km s$^{-1}$ where the co-efficients in Equation 1 are the adopted components of the solar velocity with respect to the LSR \citep{1981gask.book.....M,2006A&A...445..545P}, and the co-efficient in Equation 2 is the adopted Galactic rotational velocity \citep{2008gady.book.....B}. The resulting v$_{GSR}$ was normalised by dividing by $\cos(b)$. This normalisation results in a "planar RV" (i.e. the observed RV on the Galactic equator), which projects the velocities of the disk stars into a more compact distribution, facilitating the easier detection of Óoutlier streamsÓ from the bulk motion of the disk stars. 

Fig.~\ref{fig:RVDF_stream} shows v$_{GSR}$/$\cos(b)$ as a function of Galactic longitude and v$_{GSR}$ as a function of latitude around NGC 3201, where background stars are shown in black dots, and in red symbols we highlight the location of NGC 3201 and candidates of tidal debris. Stars with very high RVs are relatively rare \citep{2015MNRAS.447.2046H}. We selected candidates of tidal debris by searching around a locus in v$_{GSR}$/cos(b). Of these, 12 members within the tidal radius of NGC 3201 were reported in \citet{2015MNRAS.451.1229A} shown in red triangles, and 2 stars outside the tidal radius were reported in \citet{2014A&A...572A..30K}, shown in orange triangles. Here we identify an additional 9 candidates further to those already reported. We use open squares to represent the stars close to l = 276$^{\circ}$, red dots to represent the candidates closer to the cluster and open circles to represent the two stars around l = 278.6$^{\circ}$. The black dots are the other stars in the RAVE--DR4 sample that are not related to the cluster or stream. The potential tidal debris is well collimated ranging from -70$^{\circ}$ to +40$^{\circ}$ in Galactic latitude, in a longitude width of 4$^{\circ}$. Table 1 lists all the stars analysed in this study.

\begin{figure*}
  \centering  
  \includegraphics[width=1.04\columnwidth]{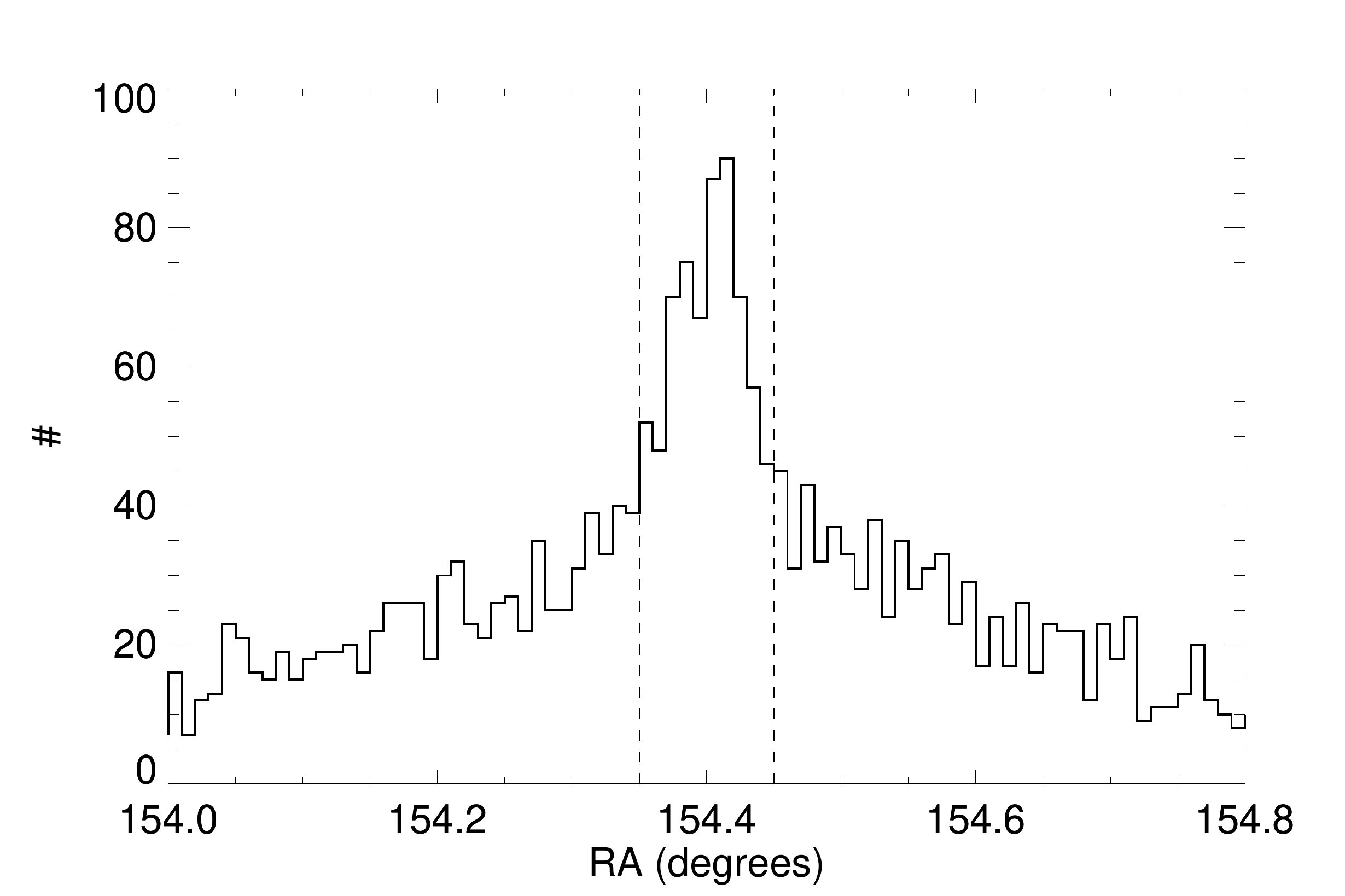}
  \includegraphics[width=1.04\columnwidth]{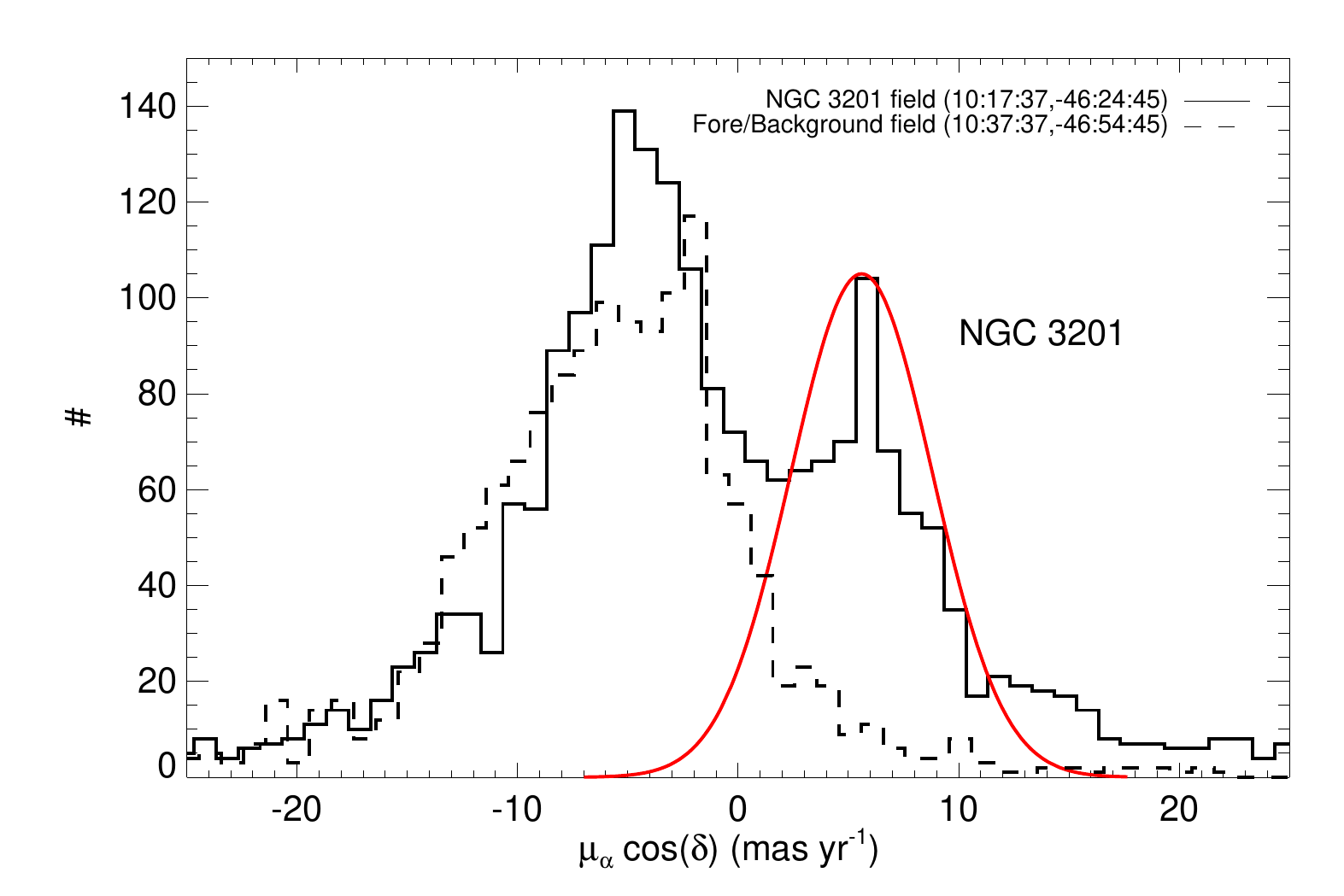}
   \caption{Left: Right Ascension (RA) distribution for stars within a radius of 0.3$^{\circ}$ centered at NGC 3201. As expected, the peak of the distribution represents the cluster. The vertical dashed lines indicate where the number density drops, at 0.05 degrees from the cluster center. Right: The proper motion distribution in RA for the 0.3$^{\circ}$ field centered at NGC 3201 (solid line). The dashed line shows the proper motion distribution for a background-only field of 0.3$^{\circ}$ radius located away from the cluster. The red gaussian distribution with a FWHM of 3.2 mas highlight the stars with a proper motion associated to NGC 3201.}
  \label{fig:RA_PM_hist}
\end{figure*}   

\begin{figure*}
  \centering  
  \includegraphics[width=2.1\columnwidth]{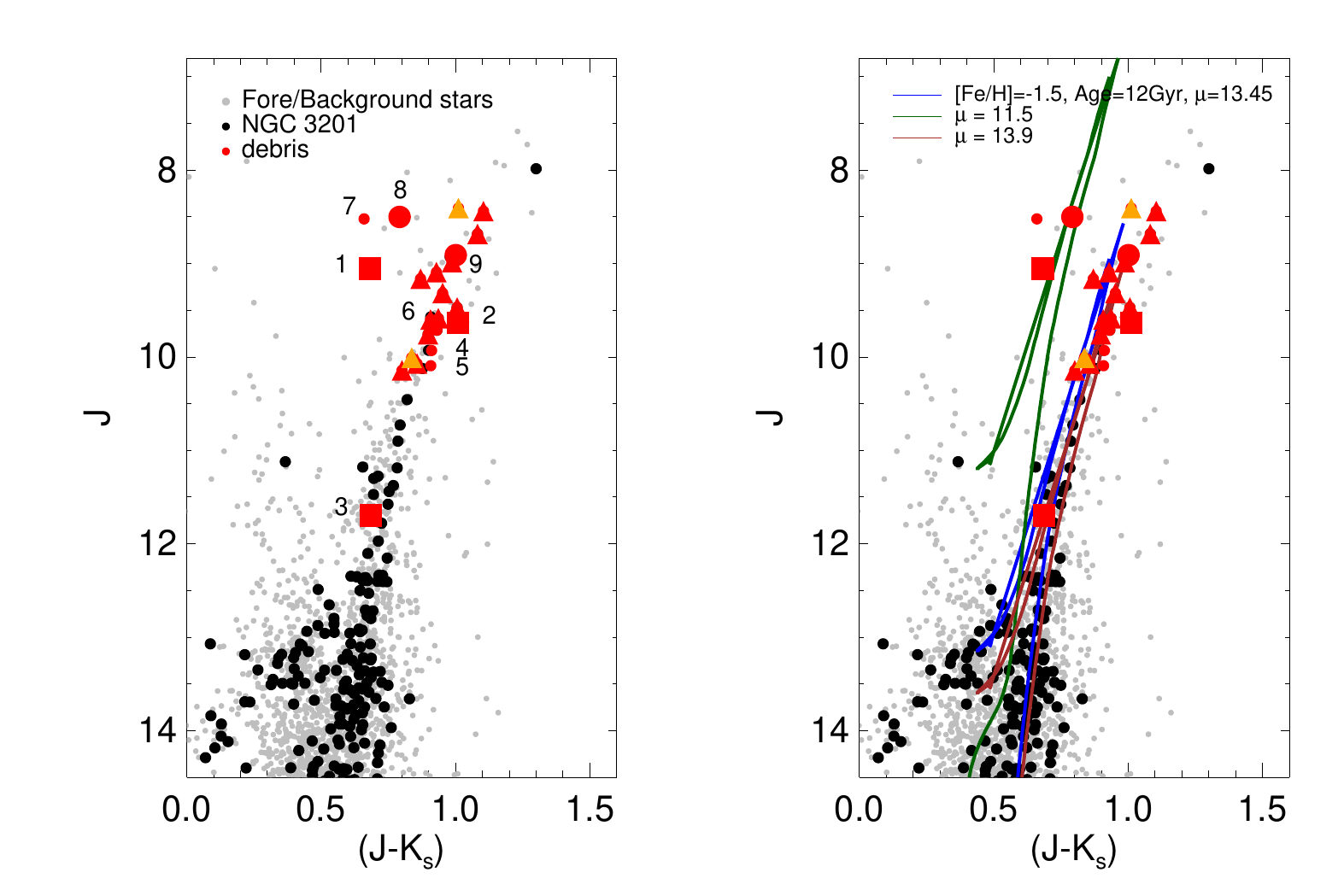}
   \caption{Color-magnitude diagram using the 2MASS J and K$_{s}$ bands for members of NGC 3201 that satisfy the proper motion selection (filled circles), possible background stars within 0.3 deg of the cluster centre (gray dots) and the tidal debris candidates (red symbols as in Fig.~\ref{fig:RVDF_stream}). In green, blue and brown a Padova isochrone of the metallicity and age reported for NGC 3201 is plotted, shifted to distances of 2, 4.9, 6 kpc respectively, and corrected for extinction.}
  \label{fig:CMD_RPD}
\end{figure*}  

\section{Proper Motions, Color-Magnitude and Metallicity Distributions}
\label{sec_ppm}

\subsection{Colour-Magntiude Diagram}

We use 2MASS bands J and K$_{s}$ \citep{2006AJ....131.1163S} to explore the photometric properties of the candidate tidal debris. From UCAC4 \citep{2013AJ....145...44Z} we selected targets within 0.3$^{\circ}$ centered in NGC 3201. We chose 0.3 degrees around the cluster in order to sample the region of sky around the cluster, including cluster members as well as a field sample in the direction of the cluster. Fig.~\ref{fig:RA_PM_hist} shows the RA and proper motion distribution for the selected field. The RA distribution in the left panel of Fig.~\ref{fig:RA_PM_hist} shows the cluster stars can be identified to be within 0.05 degrees, after which the number density sharply drops, shown in vertical dashed lines. In order to further separate the cluster and background stars we compare the PPM distribution against the PPM distribution of another 0.3$^{\circ}$ field well outside the cluster. This background-only field, located at RA: 10:37:37, DEC: -46:54:45, and its PPM distribution is plotted with a dashed line in the right panel of Fig.~\ref{fig:RA_PM_hist}. The solid black line shows the PPM distribution of the 0.3$^{\circ}$ field centered on the cluster. A clear peak is seen in this field that represent the cluster stars. The red Gaussian of a FWHM of 3.2 mas highlights the likely cluster stars. To build the CMD of NGC 3201, we selected stars that lie within in the 0.05$^{\circ}$ radius in the RA distribution as well as those that lie inside the red Gaussian in the PPM distribution. Stars that satisfy both these criteria are plotted as black filled circles in Fig~\ref{fig:CMD_RPD}. All other stars within the 0.3$^{\circ}$ field are plotted as smaller grey dots. The candidates of the tidal debris are overplotted and numbered as per Fig.~\ref{fig:RVDF_stream} and Table 1. The reddening towards NGC 3201 is E(B-V) = 0.24 mag \citep{1996AJ....112.1487H}. From the tabulated extinction law table in \citet{1990ARA&A..28...37M} we use A(J)/A(V) = 0.282, hence E(J-K$_{s}$) will be of the order of $\sim$ 0.07 mag which does not affect the CMD significantly. Therefore we do not apply any reddening corrections. 

The isochrones in Fig.~\ref{fig:CMD_RPD} were selected from \citet{2000A&AS..141..371G} for  [Fe/H] = $-1.5$ and age = 12 Gyr to match the metallicity and age reported for this cluster \citep{1996AJ....112.1487H, 2009ApJ...694.1498M}. The Red Giant Branch (RGB) of NGC 3201 appears clearly in the CMD. We observed that most of the debris stars follow the RGB. The candidates are in good agreement with the selected isochrones of cluster age and metallicity, with any deviation from the nominal RGB depending on the distance from the cluster (see Section 4).

\subsection{Proper Motion Diagram}

\begin{figure*}
  \centering  
  \includegraphics[width=1.9\columnwidth]{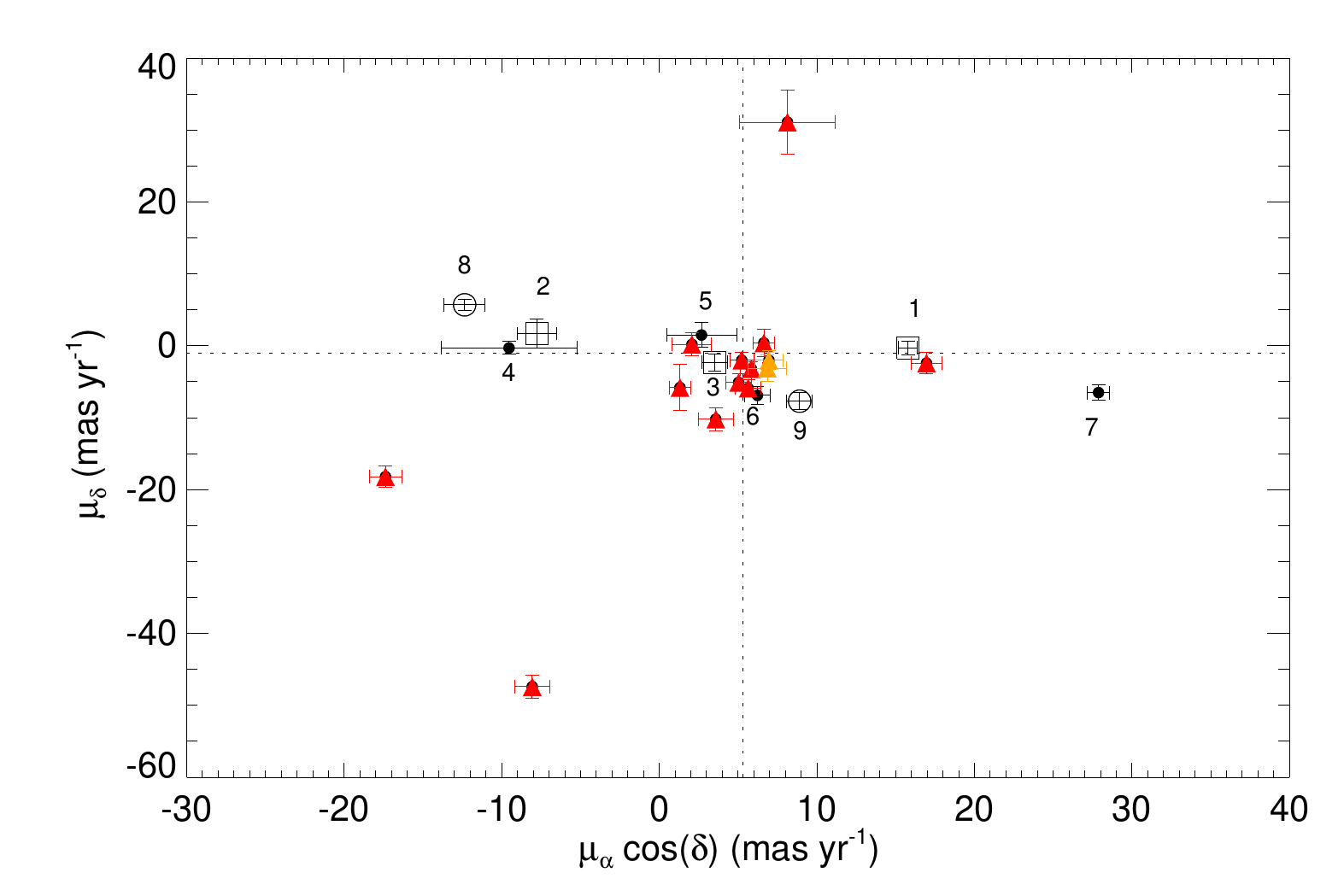}
   \caption{Proper motion diagram (mas yr$^{-1}$) for the stars selected as members of the extended tidal stream of NGC 3201. Symbols are same as in Fig.~\ref{fig:RVDF_stream}. The dashed lines are the nominal proper motion values reported for this cluster.}
  \label{fig:PM_stream}
\end{figure*}

We examine the proper motions of the tidal debris candidate using UCAC4. Fig.~\ref{fig:PM_stream} shows the proper motion diagram for NGC 3201 cluster and debris stars selected for analysis as described in Section 2. The dashed lines show the nominal proper motion values for NGC 3201 at $\mu_{\alpha}$ cos($\delta$) = 5.28 $\pm$ 0.32 mas yr$^{-1}$, $\mu_{\delta}$ = -0.98 $\pm$ 0.33 mas yr$^{-1}$ \citep{2007AJ....134..195C}. We do not expect the proper motions to be the same for all stars in an extended tidal stream. For debris stars that lie in the same part of the sky as NGC 3201 we can expect similar proper motions as the GC itself, however that will not hold for stream stars moving on the same orbit but located in a very different direction from NGC 3201. There are 3 stars with high proper motions lying inside the tidal radius of the cluster and with a similar RV reported for NGC 3201 ($\mu_{\delta}$ = -47.4 , -18.2 and 31.1 mas yr$^{-1}$ respectively). The spread in proper motions for stars inside the tidal radius of NGC 3201 is likely due to larger uncertainties in crowded fields.

\subsection{Metallicity Distribution Function}
\label{sec_mdf}

\begin{figure}
  \centering  
 \includegraphics[width=1.03\columnwidth]{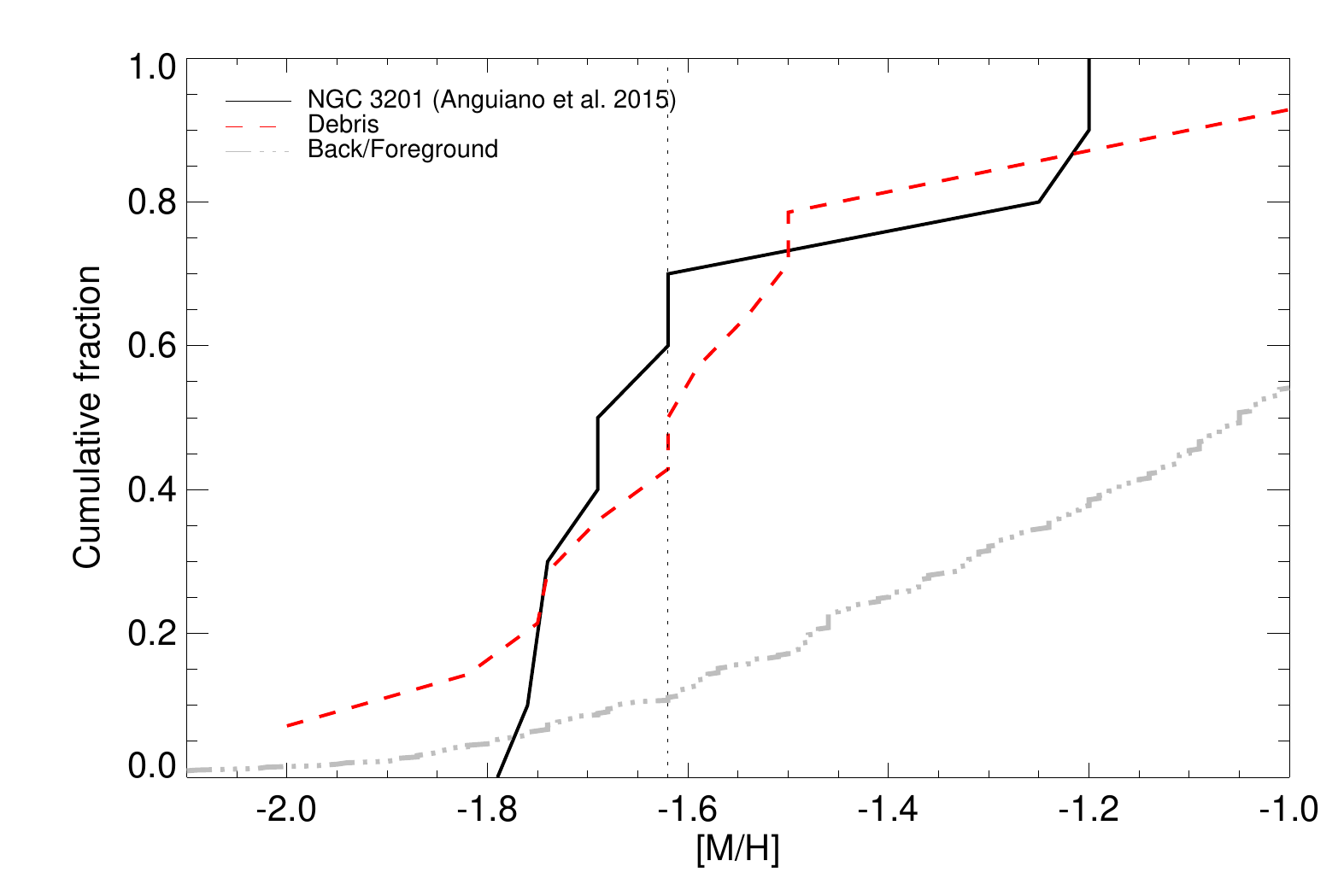}
   \caption{Metallicity cumulative fraction for the tidal debris associated with NGC 3201 (red dashed line) and for NGC 3201 (black line) using only the stars reported in \citet{2015MNRAS.451.1229A}. A back/foreground field is shown in grey dashed line.}
  \label{fig:MDF}
\end{figure}

We also explore the RAVE-DR4 metallicities. Conservatively the metallicities are accurate to $\sim$ 0.2 dex. We find the potential tidal debris (not including the cluster stars at l= 277.2$^{\circ}$ depicted in red triangles in Fig.~\ref{fig:RVDF_stream}) to range in metallicity from  -1.2 to -1.9 dex, with the MDF peak around -1.7 dex. The black line in Fig.~\ref{fig:MDF} shows the cumulative fraction for cluster members identified in RAVE by \citet{2015MNRAS.451.1229A}, the dashed red line the fraction for the debris candidates and the gray-colored line a back/foreground field around the cluster. A KS-test shows that the maximum difference in cumulative fraction is D = 0.2. Given the limited accuracy of the RAVE low-resolution spectra and small number statistics of this sample, it is difficult to assess if the spread in metallicity seen in the MDF is real. 

\section{Space velocities and integrals of motion space}
\subsection{Distances and Space velocities}
\begin{figure*}
  \centering  
  \includegraphics[width=1.72\columnwidth]{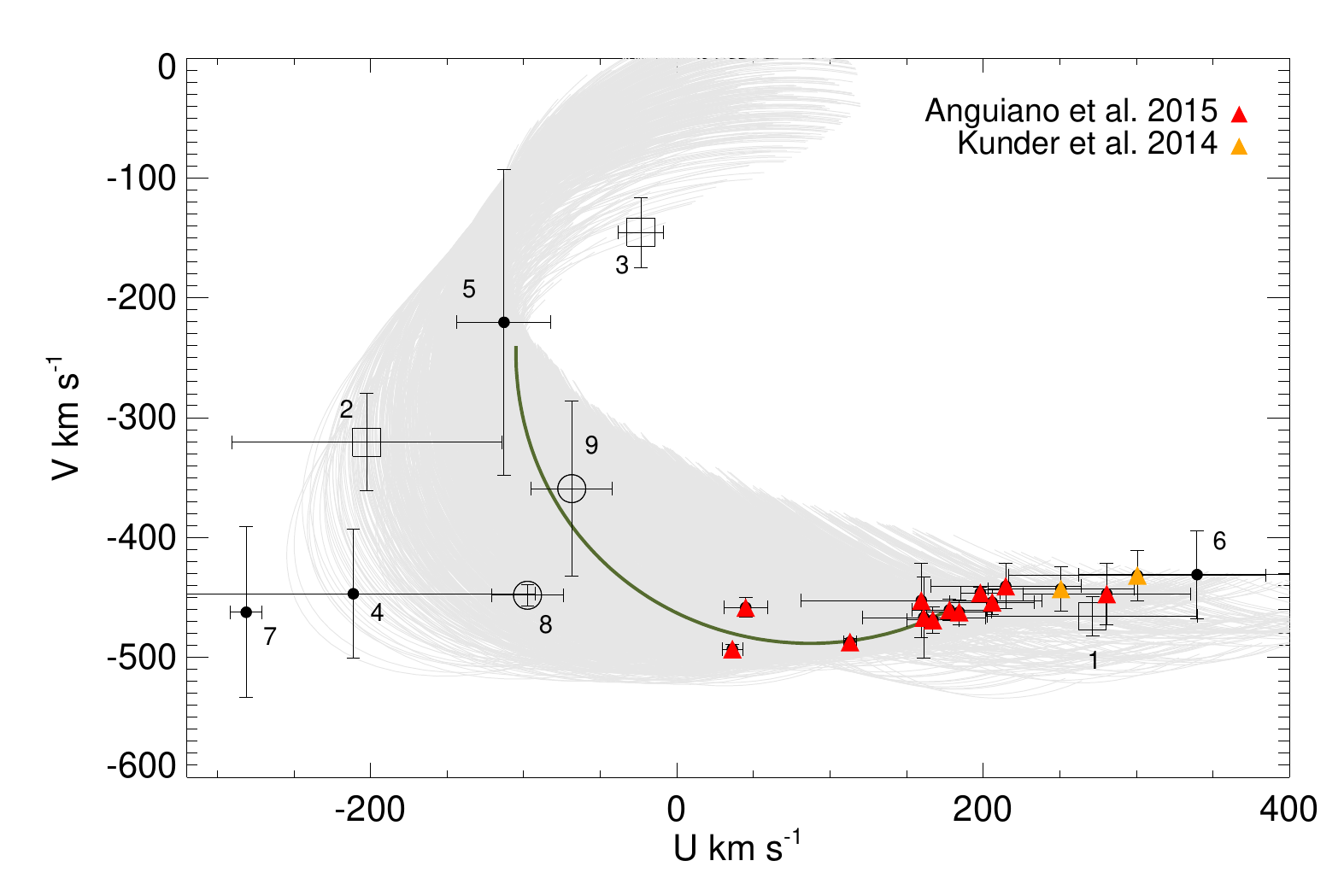}
   \includegraphics[width=1.72\columnwidth]{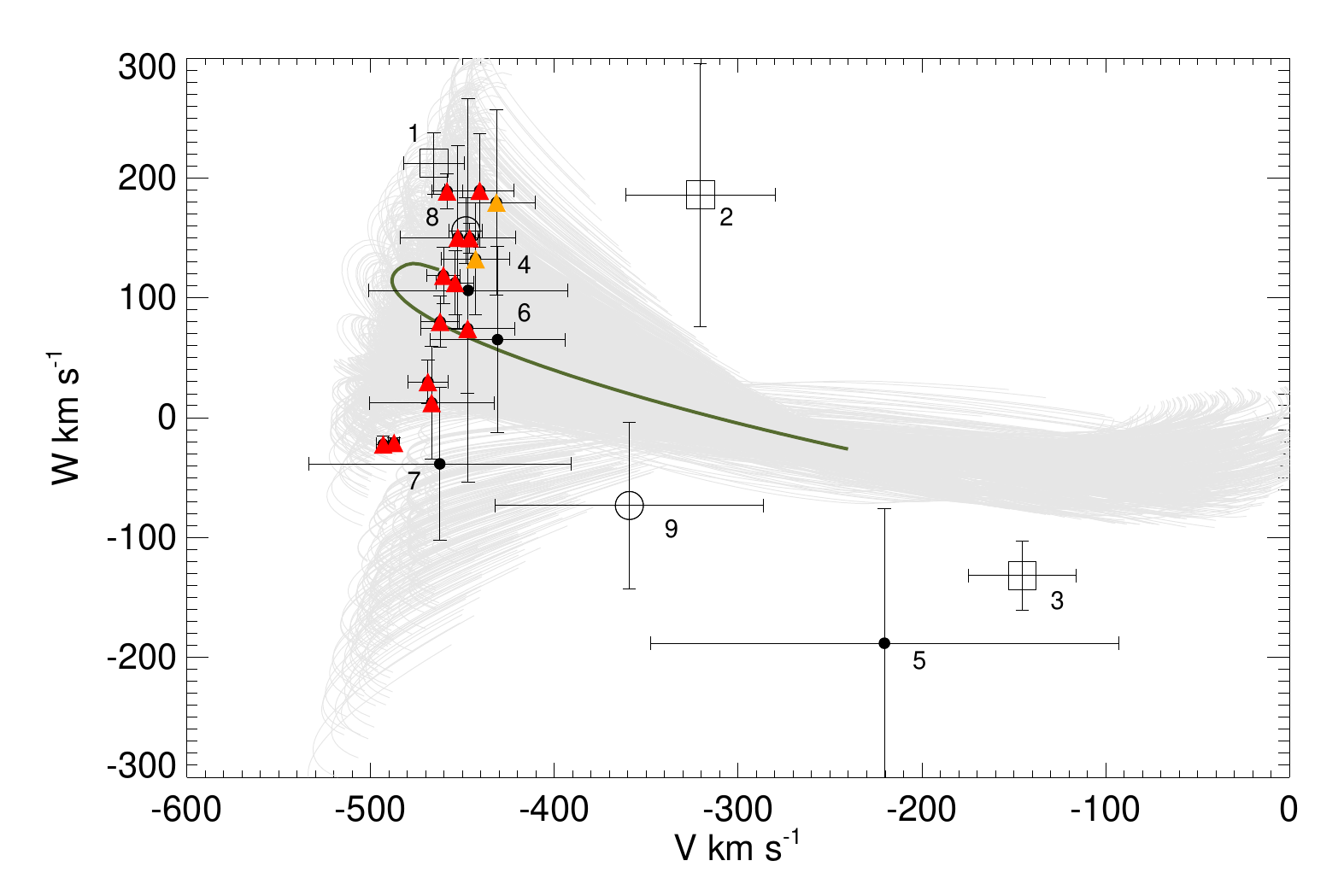}   
    \caption{U-V, V-W diagrams for the cluster and tidal debris candidates (symbols are as in Fig.~\ref{fig:RVDF_stream}). The gray area shows all the possible orbits computed using the mean values in RV, proper motions and DR5 distances reported for cluster members (red triangles) and 1-sigma variations. The green line is the orbit computed using the nominal RV, proper motion and distance values reported for NGC 3201 in the literature, d = 4.9 kpc and $\mu_{\alpha}$ cos($\delta$) = 5.28 $\pm$ 0.32 mas yr$^{-1}$, $\mu_{\delta}$ = -0.98 $\pm$ 0.33 mas yr$^{-1}$ \citep{1996AJ....112.1487H, 2007AJ....134..195C}.}
  \label{fig:UVW}
\end{figure*}

Velocities in a Cartesian Galactic system in the direction of the Galactic centre and rotation at the Sun were obtained following the method developed by \citet{1987AJ.....93..864J}. The velocities are defined in a right-handed Galactic system with U pointing towards the Galactic centre, V in the direction of rotation, and W towards the north Galactic pole. We used the new distances derived by  \citet{2014MNRAS.437..351B} for the forthcoming 5th Data Release (Kunder/Steinmetz priv. com.) For 9 stars of our sample there are no RAVE distances available. As 9 of these stars have Galactic coordinates close to NGC 3201, we use a simple reduced proper motion analysis to estimate their distances. We have the distance modulus

\begin{equation}
\mu_{J} = J - H_{J} + 5 \log v_{T} - 3.379 
\end{equation}

where v$_{T}$ is the tangential velocity. We use the v$_{T}$ of NGC 3201 (v$_{T}$ = 124 km s$^{-1}$) to compute the stellar distances for the 9 stars that have Galactic coordinates similar to NGC 3201 and also similar RV. A change in the cluster distance of 1 kpc gives a variation in v$_{T}$ of 22 km s$^{-1}$. We use this value to compute the errors in the distances. We also perform a reddening correction in the J band using A(J) = 0.216 \citep{1990ARA&A..28...37M}. The distances derived via the reduced proper motion analysis are consistent with the distances derived by \citet{2014MNRAS.437..351B} where the mean difference is 1.7 $\pm$ 0.8 kpc for the 5 stars in common. In both cases, we find the tidal debris candidates are mostly located at distances ranging from 1 to 7 kpc with a peak around 4 kpc. The reported distance for NGC 3201 is 4.9 $\pm$ 0.3 kpc \citep{1996AJ....112.1487H}. Fig.~\ref{fig:UVW} shows the locations of the cluster and debris stars in the U,V and V,W plane.

The cluster members clump together in the phase space, V (-460 km s$^{-1}$), but show a spread of $\sim$ 200 km s$^{-1}$ in the U and W velocities.  This is most likely due to uncertainties in the proper motion (see also \citealt{2015MNRAS.451.1229A}), and therefore although the phase space can help decipher space velocity correlations within the tidal debris candidates, we caution against imposing a strict criteria in the U, V, W plane to select tidal debris members. These objects reported in \citet{2014A&A...572A..30K, 2015MNRAS.451.1229A} are highlighted as red/orange triangles. The debris candidates have velocities ranging from +120 to -180 km s$^{-1}$ in U, while in the V component they range from -500 to -100 km s$^{-1}$  and in W we find values from -220 to +250 km s$^{-1}$. In the U-V plane, the extended candidates appear to form an arc-like structure going from negative to positive values in U, while the V velocity approaches the velocity of the cluster members. In the V-W plane, is clear that even the member stars show a spread in W velocity. In principle, stars inside a GC tidal radius should follow similar velocities. For the stars inside the tidal radius we re-computed the velocities using a nominal distance for these stars of 4.9 kpc. We found that the spread is even larger; suggesting problems with proper motions in crowded fields. We used GalPy \citep{2015ApJS..216...29B} to calculate orbits backward in time over 6 Gyr. The gray area in Fig.~\ref{fig:UVW} shows all the possible orbits computed using the mean values in RV, proper motions and RAVE-DR5 distances reported for cluster members (red triangles) and 1-sigma variations. We found a large spread in orbits. In the U-V plane most of the tidal debris candidates are inside the area drawn by the orbits while in the V-W plane, 5 tidal debris stars are within the region of the orbits, and 3 candidates with negative W values lie close to (but not inside) the regions predicted by the orbits.

\begin{figure*}
  \centering  
    \includegraphics[width=1.9\columnwidth]{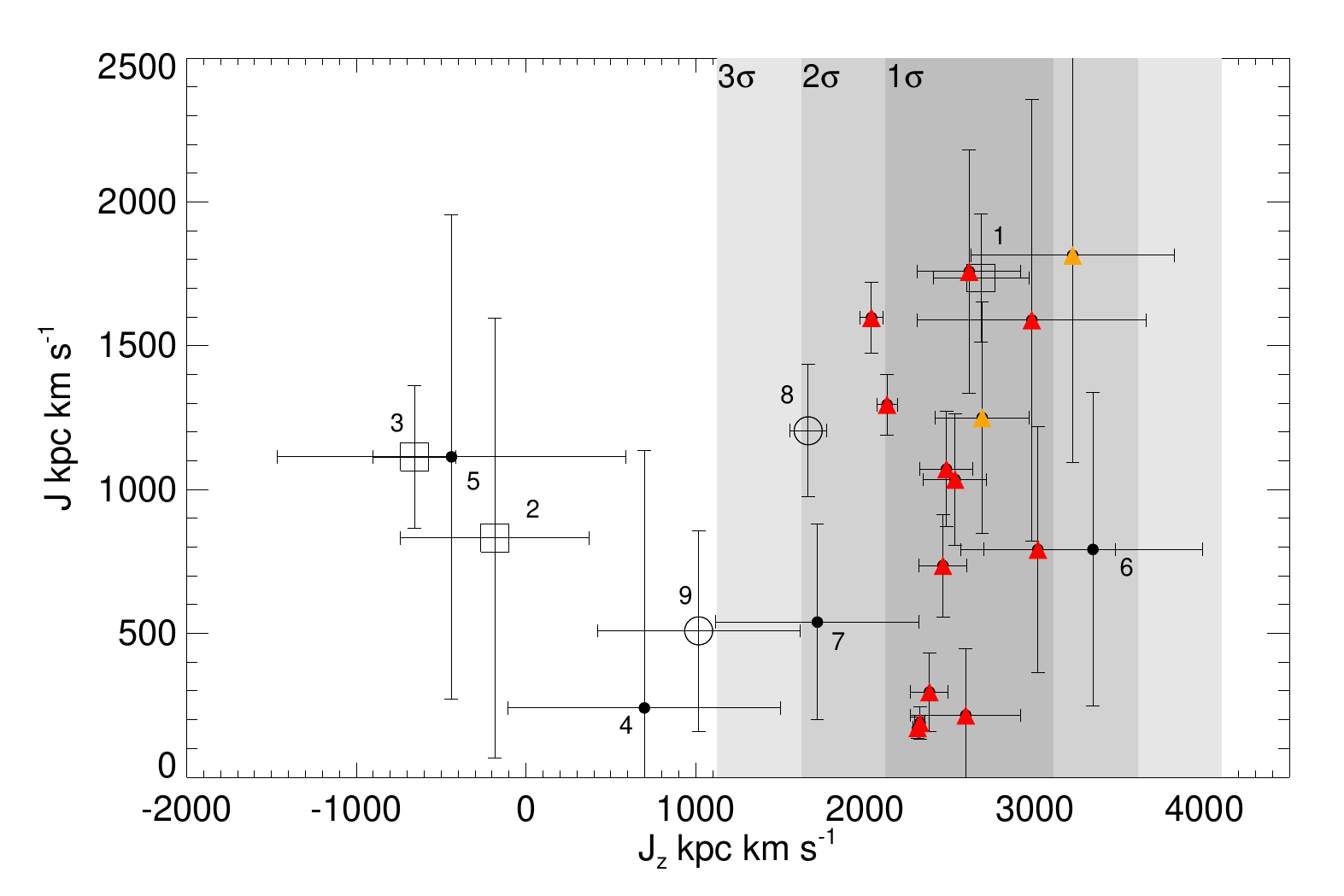}  
   \caption{(J,J$_{z}$) space for the tidal debris candidates associated with NGC 3201 (symbols as in Fig.~\ref{fig:RVDF_stream}). J here is (J$_{x}$$^{2}$ + J$_{y}$$^{2}$)$^{1/2}$. J$_{z}$ $>$ 0 in our convention means retrograde. To computed the non-independent random errors in the angular momentum we used a Monte Carlo evaluation. We found a larger spread in J$_{z}$ for cluster members inside the tidal radius. The nominal value for NGC 3201 is (J,J$_{z}$) $\sim$ (1200,2700) kpc km s$^{-1}$. The gray shaded areas are the 68, 95 and 99$\%$ confidence limit respectively.}
  \label{fig:AM}
\end{figure*} 

\subsection{Angular Momentum}

Fig.~\ref{fig:AM} shows the (J,J$_{z}$) space for the cluster and debris stars associated with NGC 3201. To propagate the non-independent random errors in the angular momentum we used a Monte Carlo evaluation taking into account the uncertainties in distances and space velocities. If these are conserved quantities, the debris candidates would share the same orbital plane. In an axisymmetric potential, J$_{z}$ should be a constant value for the tidal debris candidates when reliable and precise distances and proper motions are available. As expected from the significant spread in the space motions we found that the stars inside the tidal radius of the cluster (red triangles) have a $\Delta$J$_{z}$ $\sim$ 1000 kpc km s$^{-1}$. The large spread in angular momentum for a low mass cluster like NGC 3201 reiterate problems with proper motions or distances for these stars. We run a Monte Carlo test to compute the scatter in J$_{z}$ for the cluster members (red triangle). The shades of gray areas in Fig.~\ref{fig:AM} shows the 68, 95 and 99$\%$ confidence limit. We found 4 debris candidates have similar angular momentum values to the cluster to within the errors. One large square, one large circle and 2 dots (see Fig.~\ref{fig:AM}) are within 2-sigma area. Note that in our adopted right-handed convention, positive J$_{z}$ values represent a retrograde orbit. In this framework a typical Milky Way population will have negative J$_{z}$ value and will lie on the left side of Fig.~\ref{fig:AM}.


\begin{table*}
 \centering
 \begin{minipage}{160mm}
  \caption{Properties of the potential tidal debris candidates (squares, dots, circles) and cluster members (triangles). Symbols and the ID numbers are as in figures.}
  \begin{tabular}{@{}llrrcrrrlrrr@{}}
  \hline
   ID -- Symbol  & l ($^{\circ}$) & b ($^{\circ}$) & J & J - K$_{s}$ & $\mu_{\alpha}*cos(\delta)$ & $\mu_{\delta}$ & U & V & W & [M/H] & Dist  \\
   & & &  &  & (mas/yr) & (mas/yr) &  & (km/s) & & (dex) & (kpc) 
  \\
 \hline
1 -- Square & 275.5&-12.1&  9.05&  0.68& 15.8$\pm$0.6 & -0.3$\pm$1.0 &  +271 & -465 & +212 & -2.10 & 2.5$\pm$0.7  \\
2 -- Square &275.7& 39.6&  9.63&  1.01& -7.8$\pm$1.2 &  1.7$\pm$2.0 & -202 & -320 & +186 & -1.50 &  6.4$\pm$0.9 \\
3 -- Square  &275.9&-67.5& 11.69&  0.69 &  3.5$\pm$0.8& -2.3$\pm$1.2 &  -23 & -145 & -131 & -1.00 & 3.8$\pm$1.2   \\
4 -- Dot &276.6& 31.9&  9.93&  0.91 & -9.5$\pm$4.3& -0.3$\pm$0.9 & -211 & -446 & +106 & -1.60 & 6.2$\pm$1.4 \\
Orange triangle&276.8&  8.2& 10.00&  0.84&  7.0$\pm$0.9& -2.0$\pm$1.3 & +300 & -431 & +179 & -1.82 & 5.2$\pm$1.4\\
5 -- Dot &276.9&-56.5& 10.09&  0.91&  2.7$\pm$2.2&  1.5$\pm$1.7& -112 & -220 & -188 & -1.50 & 7.1$\pm$1.5  \\
Red triangle&277.1&  8.7&  9.16&  0.87&  6.6$\pm$0.7&  0.4$\pm$1.9 & +214 & -440 & +189 & -1.75 & 4.0$\pm$1.0 \\
6 -- Dot &277.2&  8.8&  9.71&  0.93&  6.2$\pm$0.8& -6.9$\pm$1.2 & +339 & -430 & +65 & -1.80 & 5.0$\pm$1.4    \\
Red triangle&277.2&  8.4& 10.14&  0.80&  5.6$\pm$0.8& -5.9$\pm$1.2& +280 & -447 & +74 & -1.80 & 4.4$\pm$1.3  \\
Red triangle&277.2&  8.7&  9.59&  0.91&-17.4$\pm$1.0&-18.2$\pm$1.5& +36 & -493 & -22 & -1.70 & 0.8$\pm$0.2 \\
Red triangle&277.2&  8.6&  8.68&  1.08&  1.3$\pm$0.7& -5.8$\pm$3.2&  +161 & -466 & +12 & -1.70 & 3.9$\pm$0.8 \\
Red triangle&277.2&  8.6&  9.31&  0.95& -8.1$\pm$1.1&-47.4$\pm$1.6& +113 & -487 & -21 & -1.62 & 0.7$\pm$0.1 \\
Red triangle&277.2&  8.6& 10.07&  0.86&  8.1$\pm$3.0& 31.1$\pm$4.4& +44 & -458 & +188 & -1.20 & 0.8$\pm$0.2 \\
Red triangle&277.2&  8.7&  9.47&  1.00&  2.1$\pm$1.2&  0.2$\pm$1.6& +159 & -452 & +150 & -1.76 & 7.8$\pm$1.4 \\
Red triangle&277.2&  8.6&  9.09&  0.93& 17.0$\pm$1.0& -2.4$\pm$1.5&  +198 & -446 & +149 & -1.62 & 1.2$\pm$0.3 \\
Red triangle&277.3&  8.6&  8.98&  0.99&  5.2$\pm$0.8& -2.0$\pm$1.1&  +177 & -460 & +118 & -1.35 & 3.0$\pm$0.8 \\
Red triangle&277.3&  8.6&  9.58&  0.94&  3.6$\pm$1.1&-10.2$\pm$1.6&  +167 & -468 & +29 & -0.80 & 2.0$\pm$0.4 \\
Red triangle&277.3&  8.6&  9.75&  0.89&  5.0$\pm$0.8& -5.1$\pm$1.2&  +184 & -462 & +80 & -1.74 & 2.6$\pm$0.6 \\
7 -- Dot &277.4&-62.3& 8.52&  0.66& 27.9$\pm$0.7& -6.5$\pm$1.1&  -281 & -462 & -38 & -2.00 & 2.4$\pm$0.6 \\
Red triangle&277.5&  8.6&  8.44&  1.10&  5.8$\pm$0.7& -3.2$\pm$1.0&  +205 & -453 & +112 & -1.25 & 3.1$\pm$0.8 \\
Orange triangle&277.8&  8.6&  8.41&  1.01&  6.9$\pm$1.2& -3.1$\pm$1.8&  +250 & -442 & +132 & -1.69 & 3.7$\pm$0.7\\
8 -- Circle &278.5& 21.3&  8.50&  0.79&-12.4$\pm$1.3&  5.7$\pm$0.8& -97 & -448 & +155 & -1.54 & 2.2$\pm$0.5 \\
9 -- Circle &278.7&-62.5&  8.91&  1.00&  8.9$\pm$0.8& -7.7$\pm$1.2 & -68 & -359 &  -73  & -1.25 & 4.1$\pm$0.8\\
\hline
\end{tabular}
\end{minipage}
\end{table*}

\section{Conclusions}    

Exploring the radial velocity in the Galactic Standard of Rest frame (v$_{GSR}$) in the RAVE-DR4 catalogue we detected extended tidal debris which may be associated with the halo GC, NGC 3201. Due to the extended nature of the debris and the resulting projection effects, traditional membership indicators such as positions on the CMD or proper motions are not applicable. Therefore verifying the membership of the tidal debris in not trivial.

The space velocities and orbital motions show interesting features, the interpretation of which is however impacted by uncertainties in proper motions and distances. If we enforce that tidal debris from NGC 3201 should share common J, J$_{z}$ and should lie inside the predicted orbits in U, V, W, then we have 4 candidates, noted as a large square, large circle and 2 small dot symbols in Fig.~\ref{fig:UVW} and Fig.~\ref{fig:AM}. These same stars also match the isochrone of NGC 3201 age and metallicity displaced by a distance corresponding to 2 kpc (see Fig.~\ref{fig:CMD_RPD}). We highlight that these three stars are located at -12$^{\circ}$, +21$^{\circ}$ and -62$^{\circ}$ angular extent on sky from the cluster center, together spanning 80 degrees across the sky. 

The confirmation of tidal debris around NGC 3201 could help to understand two-body interactions in the cluster and evaporation processes, also it will be a tell-tale sign of its origins, currently speculated to be extra-Galactic in nature due to its high radial velocity and retrograde orbit. No other tidal debris of such a large spatial extent has been reported around this cluster to date. The presented sample is limited in size to reliably calculate orbits and subsequent implications on the Galactic potential. However, the brightness of the tidal debris candidates makes a detailed follow-up chemical abundance analysis via high-resolution spectroscopy relatively straight-forward. Further exploration of tidal debris around NGC 3201 is encouraged, in particular, a search for lower-mass stars, which although fainter, have been shown to preferentially reside in tidal tails \citep[e.g.][]{1999A&A...352..149C,2004AJ....128.2274K}.
    
\section*{Acknowledgments}

BA gratefully acknowledge the financial support of the Australian Research Council through Super Science Fellowship FS110200035. DZ acknowledge the support of the ARC through Future Fellowship FT110100743. We thank the Centre de Donn\'{e}es Astronomiques de Strasbourg (CDS), the U. S. Naval Observatory and NASA for the use of their electronic facilities, especially SIMBAD, ViZier and ADS. Funding for RAVE has been provided by: the Australian Astronomical Observatory; the Leibniz-Institut f\"{u}r Astrophysik Potsdam (AIP); the Australian National University; the Australian Research Council; the French National Research Agency; the German Research Foundation (SPP 1177 and SFB 881); the European Research Council (ERC-StG 240271 Galactica); the Istituto Nazionale di Astrofisica at Padova; The Johns Hopkins University; the National Science Foundation of the USA (AST-0908326); the W. M. Keck foundation; the Macquarie University; the Netherlands Research School for Astronomy; the Natural Sciences and Engineering Research Council of Canada; the Slovenian Research Agency; the Swiss National Science Foundation; the Science \& Technology Facilities Council of the UK; Opticon; Strasbourg Observatory; and the Universities of Groningen, Heidelberg and Sydney. The RAVE web site is at http://www.rave-survey.org

\appendix

\bsp

\label{lastpage}

\end{document}